\documentclass[aps,prb,twocolumn,groupedaddress,showpacs]{revtex4-1}

\usepackage{dcolumn} % Align table columns on decimal point
\usepackage{bm} % bold math
\usepackage{amsthm}
\usepackage{latexsym}
\usepackage{float}
\usepackage{amsfonts}
\usepackage{amsmath}
\usepackage{amssymb}
\usepackage{color,graphicx}
\usepackage{booktabs}
\usepackage{epstopdf}
\begin{document}

\title{Dephasing of InAs quantum dot {\it p}-shell excitons using two-dimensional coherent spectroscopy}
\author{Takeshi Suzuki$^{1, 2}$, Rohan Singh$^{1, 2}$, Galan Moody$^3$, Marc A{\ss}mann$^4$,\\
Manfred Bayer$^4$, Arne Ludwig$^5$, Andreas D. Wieck$^5$, and Steven T. Cundiff$^{1, 2}$}
\affiliation{$^1$JILA, University of Colorado \text{\&} National Institute of Standards and Technology, Boulder, Colorado 80309-0440, USA\\
$^2$Physics Department, University of Michigan, Ann Arbor, MI 48109, USA\\
$^3$National Institute of Standards and Technology, Boulder, Colorado 80305, USA\\
$^4$Experimentelle Physik 2, Technische Universt\"{a}t Dortmund, D-44221 Dortmund, Germany\\
$^5$Lehrstuhl fuer Angewandte Festkoerperphysik, Ruhr-Universitaet Bochum, Universitaetsstrasse 150, D-44780 Bochum, Germany
}
\date{\today}

\begin{abstract}
The dephasing mechanisms of {\it p}-shell and {\it s}-shell excitons in an InAs self-assembled quantum dot ensemble are examined using two-dimensional coherent spectroscopy (2DCS). 2DCS provides a comprehensive picture of how the energy level structure of dots affects the exciton dephasing rates and recombination lifetimes.
We find that at low temperatures, dephasing of {\it s}-shell excitons is lifetime limited, whereas {\it p}-shell excitons exhibit significant pure dephasing due to scattering between degenerate spin states.
At elevated temperatures, quadratic exciton-phonon coupling plays an important role in both {\it s}-shell  and {\it p}-shell exciton dephasing.
We show that multiple {\it p}-shell states are also responsible for stronger phonon dephasing for these transitions
\end{abstract}


\maketitle

\section{Introduction}
The strong light-matter coupling and lifetime-limited homogeneous linewidth of semiconductor quantum dots (QDs) make them ideally suited for solid-state photonics, opto-electronic, and quantum technologies \cite{Alex2012,Stievater2001, Kamada2001, Ramsay2010}.
While studies of the dephasing rate, $\gamma$, in semiconductor QDs have primarily focused on the ground state {\it s}-shell exciton transition\cite{Uskov2000, Krummheuer2001, Muljarov2007, Grange2009,Besombes2001, Bayer2002, Favero2003, Peter2004, Ortner2004, Favero2007,Schmitt-Rink1987, Takagahara1999}, light-matter interaction with excited exciton states is also important.
For example, from a quantum optics perspective, optical manipulation of excitons in the first excited state ({\it p}-shell) allows for coherent control of QD spins \cite{Carter2010}, while scattering between the {\it s}- and {\it p}-shell broadens $\gamma$, limiting single-photon indistinguishability\cite{Gazzano2012, Dusanowski2017}. QDs also serve as an ideal gain medium for efficient solid-state lasers \cite{Nielsen2004, Berstermann2007}.
Relaxation and scattering between QD states affect the lasing dynamics and intensity fluctuations, which are especially pronounced in QD nanolasers in which few dots contribute to gain \cite{Chow2014}.

Despite the importance of {\it p}-shell excitons in QD devices, their relaxation and dephasing dynamics are not completely understood.
In conventional photoluminescence (PL) and photoluminescence emission (PLE) measurements under non-resonant excitation conditions, efficient relaxation dynamics of excited exciton states has been revealed to occur via emission of multiple LO phonons and interaction with wetting layer continuum states \cite{Farfad1995, Toda1999, Feldmann2001}.
Measurements of the fundamental linewidths using PL or PLE, however, are challenging owing to the excitation of additional carriers that can affect the dynamics and mask the intrinsic optical response. Resonant nonlinear optical techniques such as four-wave mixing (FWM) avoid the creation of excess carriers \cite{Borri2001, Borri2007}. 
FWM techniques are typically performed by integrating the optical signal using a slow photodetector.
Consequently, time-integrated FWM provides an ensemble-averaged response, which can mask any dispersion in the optical response that arises from fluctuations in the QD size or composition.
To access a complete picture of the optical response of QD excited states, both the dephasing and recombination dynamics must be characterized with sufficient energy or spatial resolution to examine individual dots.

An alternative method that circumvents these challenges is two-dimensional coherent spectroscopy (2DCS), which unfolds the nonlinear FWM signal onto two frequency dimensions\cite{Mukamel2000, Jonas2003}.
2DCS can separate between homogeneous and inhomogeneous broadening, which makes it especially useful for obtaining single-dot-like properties in QD ensembles\cite{Siemens2010, Moody2017}.
Recently, it has also been extended to study coherent evolutions of a QD ensemble \cite{Suzuki2016, Suzuki2018} and coherent interactions between QDs \cite{Kasprzak2011, Martin2018}.
Moreover, since 2DCS is a three-pulse FWM technique, it also has the capability to measure population decay rates and dephasing rates to provide insight into size-dependent dephasing mechanisms, which is the focus of this work.

We perform 2DCS of an InAs self-assembled QD ensemble to investigate the size and temperature dependence of the dephasing and population decay rates for {\it p}-shell excitons in comparison with {\it s}-shell excitons.
At low temperatures $\leq$20 K, the {\it s}-shell exciton homogeneous lineshape is Lorentzian corresponding to a narrow ($<$ 5 $\mu$eV) zero-phonon line (ZPL), whereas {\it p}-shell exciton lineshapes deviate substantially due to the presence of phonon sidebands (PSBs).
Interestingly, we observe no significant size-dependence in physical properties regarding exciton-phonon interactions for either exciton transition, which is likely due to thermal annealing of the sample that blue-shifts and narrows the inhomogeneous distribution.
Measurements of the ZPL-to-PSB ratio versus temperature reveal that the additional dephasing channels for the {\it p}-shell excitons arise from multiple, closely degenerate electron and hole states for these transitions.
We also find that the population recombination rates for both transitions are independent of temperature up to 80 K.
This conclusion is in contrast to the dephasing rates, indicating significant pure dephasing due to quadratic coupling to acoustic phonons.
We observe stronger coupling for {\it p}-shell excitons compared to {\it s}-shell excitons due to multiple electron and hole states contributing to the {\it p}-shell transitions.

\section{sample and experiment}
%Sample information
We study InAs self-assembled QDs with GaAs barriers, consisting of 10 quantum-mechanically-isolated, epitaxially-grown layers.
The sample is thermally annealed post-growth at 900 $^{\circ}$C for 30 s, which blue-shifts the resonances and results in a 100 meV in-plane confinement due to diffusion of gallium and indium atoms\cite{Sokolov2016}.
The sample was unavoidably doped during growth and annealing process, resulting in approximately half of the QDs being charged with a hole, which forms a charged exciton (trion) with a photo excited electron and hole pair \cite{Moody2013_1}. From the energy shift of the annealed sample relative to an unannealed sample (not shown), the maximum indium content is estimated to be approximately 40\% in our sample.
The shape of QDs is {\it plano-convex lens-like}, where the diameter at the bottom is around 30 nm while the height is around 5 nm\cite{Blockland2009}.
After rapid thermal annealing, this changes to a more biconvex shape due to indium outdiffusion.

\begin{figure}[!t]
\includegraphics{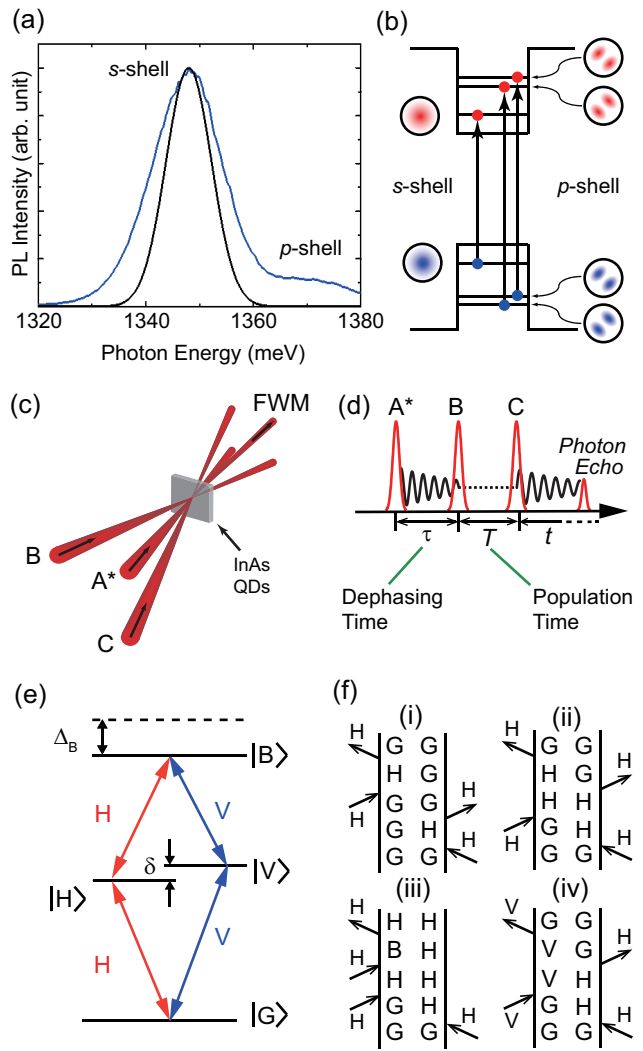}
\caption{(Color online) (a) Photo luminescence (PL) intensity from the InAs self assembled quantum dots shown as a blue line.
The excitation laser is shown as a black line.
(b) The illustration for transition in {\it s}-shell and {\it p}-shell.
%{\it s}-shell exciton is composed between the lowest states of electron and hole, whereas {\it p}-shell exciton is between the second-lowest electron and second-highest hole states.
An illustration of the wave functions for electrons and holes is also depicted. (c) Schematic and (d) time-ordering in the 2DCS experiment. Delays between A$^{\ast}$, B, C, and signal are denoted as $\tau$, $T$, and $t$, respectively. (e) Energy structure of exciton-biexciton system and selection rules. (f) Quantum paths for (i)-(iii) co-linear ({\it HHHH}) and (iv) cross-linear ({\it HVHV}) polarization sequence.}
\label{fig1}
\end{figure}

\begin{figure*}[!t]
\includegraphics{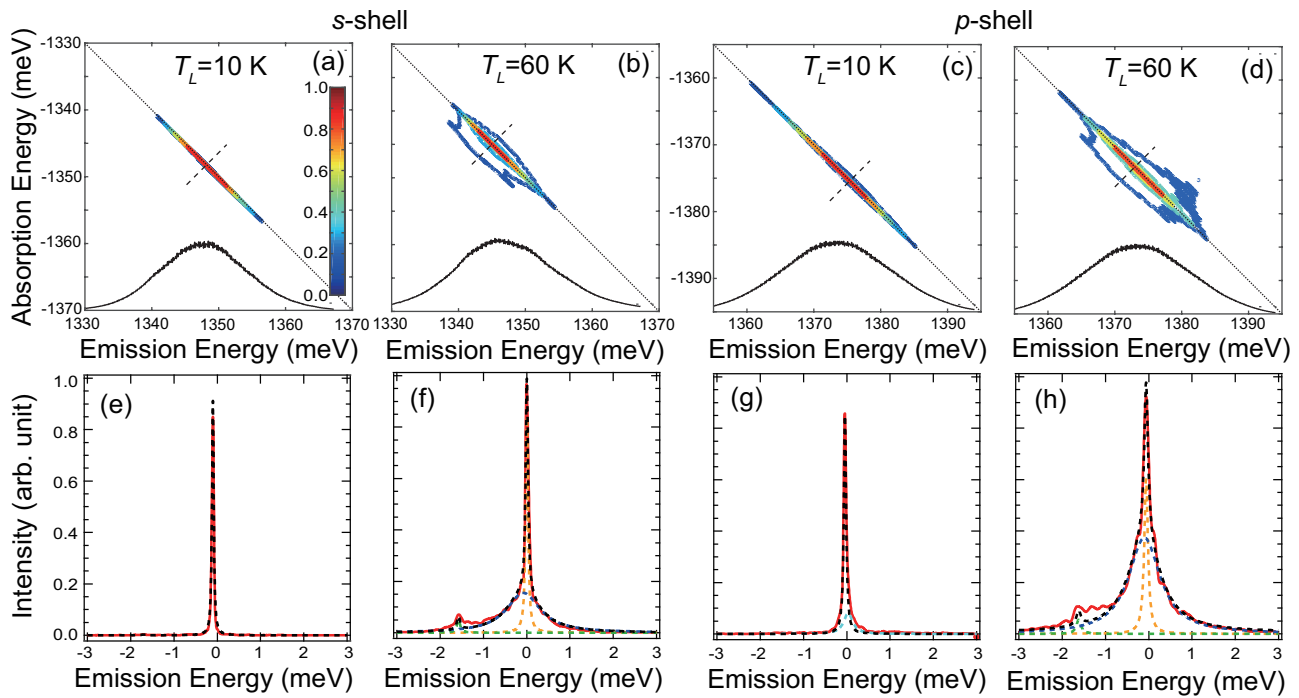}
\caption{(Color online) Normalized 2D rephasing amplitude spectra for {\it s}-shell at (a) $T_L$ = 10 K and (b) $T_L$ = 60 K and for {\it p}-shell at (c) $T_L$ = 10 K and (d) $T_L$ = 60 K.
The excitation spectrum is depicted by the black solid line in each panel. (e)-(h) Intensity profiles of cross diagonal lines shown as dashed black lines in (a)-(d). Data are shown as red circles and total fits are shown as black lines.
Lorentzian fits corresponding to zero-phonon lines, phonon side bands, and biexcitons are shown as blue-, orange-, and green-dashed lines , respectively, in (f) and (h).}
\label{fig2}
\end{figure*}

Figure 1(a) shows the photoluminescence spectrum from our sample as a blue solid line.
The black solid line represents the spectrum for the excitation laser.
The spectrum features two inhomogeneously broadened resonances near 1350 meV and 1370 meV.
As illustrated in Fig. 1(b), the lower-energy peak corresponds to the {\it s}-shell exciton composed of the lowest electron and highest hole states, while the higher energy peak corresponds to the {\it p}-shell excitons composed of the second-lowest electron and second-highest hole group states \cite{Williamson2000, Schliwa2007}.
Importantly, there are two electron and hole eigenstates for {\it p}-shell excitons, shown as two energetically-split lines and the wave functions illustrated in Fig. 1(b) \cite{Schliwa2007}.
These states are nearly degenerate as confirmed as multiplets observed in $\mu$PL spectra \cite{Canet-Ferrer2013}.
We show in the following section that these quasi-degenerate states for the {\it p}-shell are responsible for faster dephasing compared to the {\it s}-shell.

A schematic of the experiment is shown in Fig. 1(c).
2DCS is a three-pulse FWM experiment with active interferometric timing stabilization of the pulse delays \cite{Bristow2009}. Each pulse used in 2DCS has the same spectral profile and they are generated by a mode-locked Ti:sapphire laser at a repetition rate of 76 MHz.
The pulses have a bandwidth of 15 meV (full width at half maximum) as shown in Figs. 2(a)-2(d).
Three pulses are incident on the sample in a rephasing (photon-echo) time-ordering for which the conjugate pulse A$^{\ast}$ arrives at the sample first followed by pulses B and C.
Delays between pulses A$^{\ast}$, B, C, and the emitted FWM signal are denoted as $\tau$, $T$, and $t$, respectively (Fig. 1(d)).

A Fourier transform of the signal with respect to $\tau$ and $t$ generates a rephasing 2D spectrum, in which the cross-diagonal width represents the homogeneous dephasing rate $\gamma$ in the limit of large inhomogeneous broadening \cite{Siemens2010}.
Since the pulse A$^{\ast}$ is conjugated relative to the signal, the excitation energy is shown as negative values. Alternatively, a Fourier transform of the signal with respect to $T$ and $t$ generates a rephasing zero-quantum 2D spectrum in which the vertical width represents the homogeneous population decay rates, $\Gamma$ \cite{Moody_SSC}.
Measurements of both $\gamma$ and $\Gamma$ allow us to systematically study any additional processes that introduce pure dephasing ($\gamma^*$) that broadens the linewidth beyond the lifetime limit ($\gamma = \Gamma/2 + \gamma^*$).
All the measurements, unless noted otherwise, are carried out in the co-linear polarization configuration, where beam A$^{\ast}$, B, C and the emitted FWM are linearly polarized along the horizontal lab-frame direction.

Figure 1(e) shows the energy levels and selection rules for {\it s}-shell and {\it p}-shell excitons \cite{Holtkemper_2018}.
The two non-degenerate horizontal ($\left|H \right>$) and vertical ($\left|V \right>$) exciton states have orthogonal linearly polarized transitions from the ground-state ($\left|G \right>$) due to the electron-hole exchange interaction with broken cylindrical symmetry of the QD shape, which is common due to anisotropic strain in the crystal lattice, and non-symmetric shape due to anisotropic surface diffusion during the self-assembly process.
The anisotropic exchange interaction lifts the $\left|H \right>$ and $\left|V \right>$ degeneracy by the fine-structure splitting energy, $\delta$ \cite{Bayer2002_2}.
The energy of the biexciton state $\left| B \right>$ is lower than the sum of the two exciton energies by the biexciton binding energy, $\Delta_B$ \cite{Rodt2002}.
For {\it s}-shell excitons, $\delta$ and $\Delta_B$ are measured to be $19 \pm 1$ $\mu$eV \cite{Moody_SSC} and  $3.3 \pm 0.03$ meV \cite{Moody2013_2}, respectively.
Possible quantum paths for the polarization sequence of {\it HHHH} are shown in Fig. 1(f) (i)-(iii), while a quantum path for the polarization sequence of {\it HVHV} is shown in Fig. 1(f) (iv).

\section{results and discussions}
Figure 2(a) shows the normalized 2D rephasing amplitude spectrum for {\it s}-shell excitons at the lattice temperature ($T_L$) of 10 K. The spectrum exhibits a single elongated peak along the diagonal axis shown as black dotted line.
Due to the unintentional doping, almost half of QDs are charged, and inhomogeneously broadened trion and exciton peaks overlap in the spectrum on the diagonal.
However, we have confirmed that the maximum amplitude of diagonal peak from trions is two orders of magnitude weaker than the signal from excitons by comparing co-linear with cross-linear configurations, in which trion signal alone appears at the diagonal axis.
Therefore, the observed peak in Fig. 2(a) measured in co-linear configuration is mainly from the excitonic non-linear signal \cite{Moody2013_1}.

Since 2DCS can disentangle inhomogeneous and homogeneous widths along the diagonal and cross-diagonal axes, respectively \cite{Siemens2010}, we can obtain homogeneous widths by examining the cross-diagonal profile.
Figure 2(e) shows the homogeneous lineshape obtained from a cross-diagonal slice in 2D spectrum (indicated by the dashed black line in Fig. 2(a)).
The data are shown as red solid lines and reproduced well by the Lorentzian fit shown as a black dashed line.
The {\it p}-shell exciton lineshapes are obtained by tuning the center energy of the excitation laser (black solid lines in Figs. 2(c) and 2(d)) to 1372 meV. Figure 2(c) shows the 2D spectrum for {\it p}-shell excitons at $T_L$ = 10 K.
The cross-diagonal profile is shown in Fig. 2(g), and slight deviation from single-Lorentz fit (black dashed line) can be noticed.
The lineshape is fit using a double Lorentzian function.
The cyan dashed line, fit to the broad background, is attributed to phonon side bands as explained in detail below.

With increasing temperature, 2D spectra for both {\it s}-shell and {\it p}-shell excitons exhibit dramatic changes.
Figures 2(b) and 2(d) show 2D spectra at $T_L$ = 60 K for {\it s}-shell and {\it p}-shell excitons, respectively.
Peaks in both spectra shift to lower energies due to a thermal reduction in the band gap \cite{Ortner2005}, and their homogeneous lineshapes become significantly broader.
Cross-diagonal profiles are shown in Figs. 2(f) and 2(h) for {\it s}-shell and {\it p}-shell excitons, respectively.
Both spectra show a common feature--a single narrow central lineshape superimposed on a broad background peak.
This mixed lineshape has been frequently observed in single QD photoluminescence measurements \cite{Besombes2001, Favero2003, Peter2004, Favero2007}: The center lines are associated with the ZPLs corresponding to the exciton transition with no emission or absorption of phonons, while the broad background is associated with PSBs corresponding to the transition with emission or absorption of acoustic phonons.
Since the density of states for acoustic phonons is continuous due to the linear dispersion, PSBs form a continuous background as a result of the multiple transitions.
Additionally, small peaks are noticed at emission energy offset, from the peak, of around -1.7 meV in Figs. 2(f) and 2(h).
They are assigned to biexciton nonlinear signals corresponding to the quantum path of Fig 1(f) (iii), which becomes more evident in the cross-linear configuration \cite{Moody2013_1}.
The biexiton peaks also appear at the half value of biexciton binding energy of 3.3 meV because we project the corss-diagonal line profile of 2D signal onto emission energy.
Clear appearance of biexciton nonlinear signals at higher temperature can be explained by the spin-flip effect for excitons \cite{Paillard2001}, which results in decrease of polarization for {\it H}-excitons during the scan between the first and second pulse ($\tau$).
The biexciton nonlinear signal, on the other hand, is less affected by this effect because the biexciton polarization is created by the second and third pulse as shown in Fig. 1 (f) (iii), and this time duration, $T$, is fixed to 200 fs.
The results are shown as orange-, blue-, and green-dashed lines for ZPLs, PSBs, and biexcitons, respectively, in Figs. 2(f) and 2(h).
The totals of these three components are shown as black-dashed lines.
In the time domain, the fast decay component in time-integrated FWM signals corresponds to the PSBs while the slow one corresponds to ZPLs \cite{Borri2001}.

\begin{figure}[!t]
\includegraphics{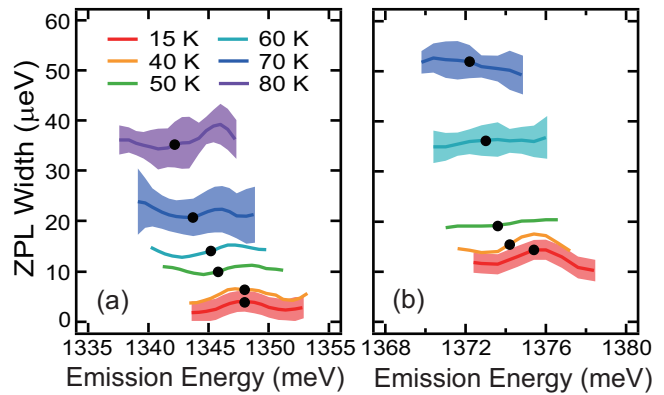}
\caption{(Color online) Zero phonon linewidths for (a) {\it s}-shell and (b) {\it p}-shell excitons across the inohomogeneous distribution as a function of temperature.
Colored regions indicate estimated errors from multiple measurements.
Black circles show the center emission energy along inhomogneous distribution at each temperature.
}
\label{fig3}
\end{figure}

\subsection{Size Dependence}
%First, we discuss the size dependence of ZPL widths.
Since the inhomogeneous broadening stems from fluctuations in the morphology of the QDs~\cite{Moody2011}, the size dependence of the ZPL widths can be inferred from their variations with emission energy.
Figures 3(a) and 3(b) show the ZPL widths corresponding to exciton dephasing rates as a function of emission energy for {\it s}-shell and {\it p}-shell excitons, respectively, for sample temperatures ranging from 15 K to 80 K.
Colored regions indicate estimated errors from multiple measurements.
Slight oscillations with respect to emission energy are a result of truncation artifacts and Fourier-transformation of the FWM data. ZPL widths for {\it s}-shell and {\it p}-shell excitons are obtained from the Lorentzian fits, in which half-width half-maximum (HWHM) corresponds to homogeneous widths $\gamma$, after deconvolution of the spectrometer response with resolution of 19 $\pm$ 2 $\mu$eV.
Regardless of temperature, no significant dependence on emission energy is found.
This is in contrast to GaAs interfacial QDs~\cite{Moody2011}, in which the ZPL widths increase for higher emission energy corresponding to smaller quantum dots size.
Similarly, no emission-energy-dependence was observed in the biexciton binding energy~\cite{Moody2013_2} and exciton homogeneous widths at the low temperature~\cite{Moody2013_1}.
These trends were attributed to the annealing effects in InAs QDs, which relieve the strain by reducing the QD/barrier lattice mismatch and minimizes built-in piezoelectric fields that can influence the optical properties.
We found that not only does annealing reduce the dispersion of the ZPLs, but it also results in a similar exciton-phonon coupling strength for all QDs in the ensemble for a given exciton shell because we have not observed size-dependence for any other physical properties in this work.
In the following, we will focus on the QD group at the center emission energy within the inhomogeneously broadened ensemble shown as black solid circles in Figs. 3(a) and 3(b).

\begin{figure}[!t]
\includegraphics{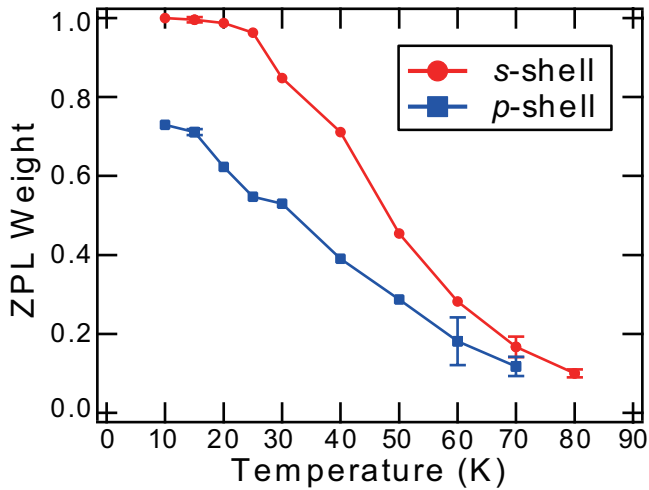}
\caption{(Color online) Zero phonon line weights for {\it s}-shell and {\it p}-shell as a function of temperature.}
\label{fig4}
\end{figure}

\subsection{Weight of zero-phonon lines}
%Next, we discuss the weight $Z$ of the ZPLs relative to the total lineshape area including the PSB background. 
We define the weight of the ZPLs, $Z$, as the ratio of areas of the fits to the ZPL and PSB.
Figure 4 shows the ZPL weight for {\it s}-shell and {\it p}-shell excitons, which both exhibit a clear decrease with increasing temperature. These behaviors are consistent with the acoustic-phonon-mediated broadening since the occupation number of acoustic phonons increases at higher temperature, which is in agreement with the previous results \cite{Borri2001, Borri2005, Cesari2010, Vagov2003}. Interestingly, the {\it p}-shell exhibits a lower $Z$ value compared to the {\it s}-shell at the lowest temperature, which is apparent from the observed PSBs in the composite lineshape in Fig 2(g).
We attribute the low-temperature PSBs for the {\it p}-shell excitons to phonon-mediated transitions between nearly degenerate electron and holes states, as observed previously in PLE measurements \cite{Htoon2001}.

\subsection{Pure dephasing for {\it s}-shell and {\it p}-shell excitons}
%Now, we discuss the pure dephasing mechanism for {\it s}-shell and {\it p}-shell excitons.
Figures 5(a) and 5(b) show rephasing zero-quantum amplitude spectra for {\it s}-shell and {\it p}-shell excitons, respectively, at $T_L$ = 10 K. 
Both spectra exhibit single peaks that are inhomogeneously broadened along the emission energy axis.
Figures 5(c) and 5(d) show the lineshape profiles along the zero-quantum energy axis in Figs. 5(a) and 5(b).
The HWHM obtained from Lorentzian fits, which are shown as black dashed lines in Figs. 5(c) and 5(d), corresponds to the population decay rates \cite{Moody_SSC}.
The population decay rate for {\it s}-shell excitons is 6 $\mu$eV corresponding to radiative lifetime of 110 picoseconds, which is in good agreement with previous work \cite{Moody_SSC}.
Surprisingly, a similar value of 6 $\mu$eV is also obtained for {\it p}-shell excitons.
A long lifetime observed in {\it p}-shell excitons has also been reported in time-resolved pump-probe transmission spectroscopy \cite{Kurze2009, Steinhoff2013}, where the fast and slow components of the lifetime were observed, and they were ascribed to the bright and dark excitons \cite{Kurtze2012}, respectively.
Our measured value of 6 $\mu$eV nearly matches the bright exciton component.
The long lifetime of both {\it s}-shell and {\it p}-shell excitons observed in this work suggests that the population transfer from {\it p}-shell to {\it s}-shell excitons is slow.
While it is difficult to determine the contribution of population transfer from wetting layer carriers or higher lying shells under non-resonant excitation conditions \cite{Kurze2009, Steinhoff2013}, our measurement under resonant excitation demonstrates that the wetting layer contribution to the population transfer is important.

\begin{figure}[!b]
\includegraphics{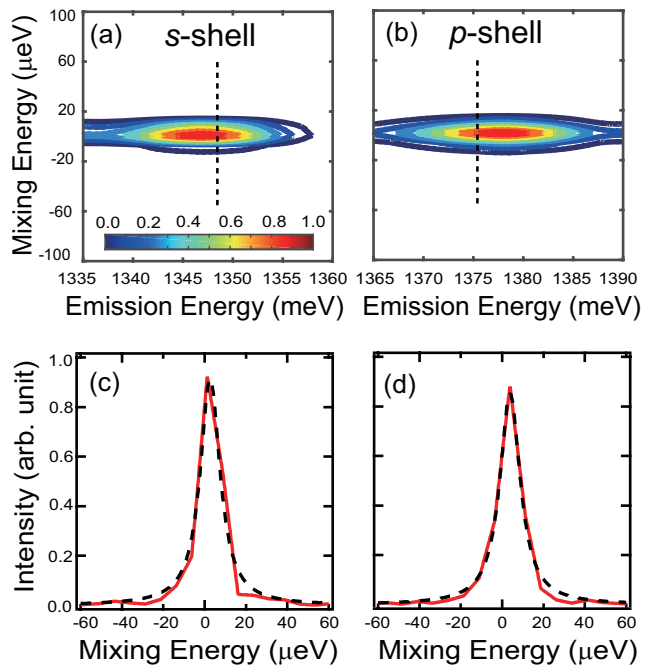}
\caption{(Color online) Rephasing zero-quantum amplitude spectra for (a) {\it s}-shell and (b) {\it p}-shell at $T_L$ = 10 K.
(c),(d) Intensity profiles of vertical lines shown as dashed lines in (a) and (b).
Red solid lines are data and black dashed lines are Lorentzian fits.
}
\label{fig5}
\end{figure}

\begin{figure}[!t]
\includegraphics{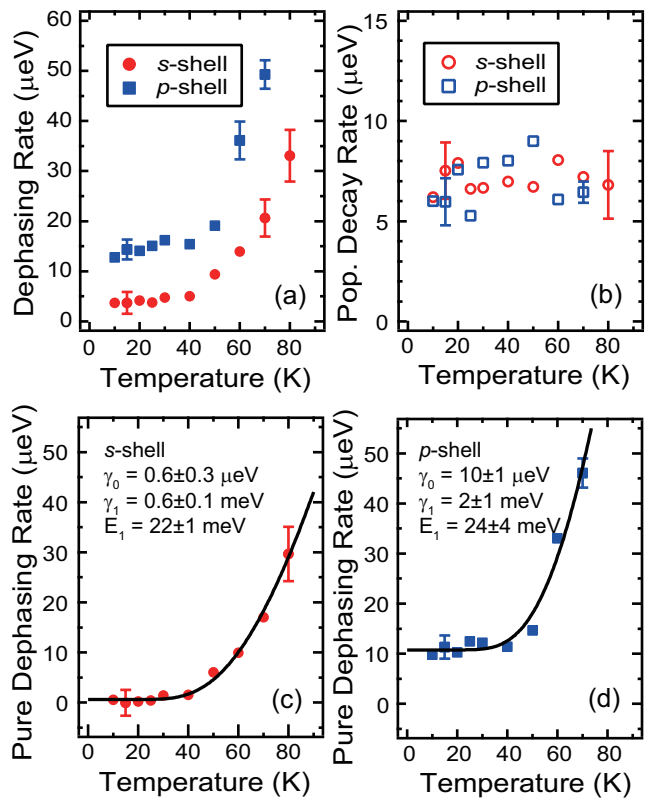}
\caption{(Color online) (a) Dephasing and (b) population decay rates for {\it s}-shell and {\it p}-shell excitons as a function of temperature.
Pure dephasing rates for (c) {\it s}-shell and (d) {\it p}-shell excitons as a function of temperature.
Marks are data and black solid lines are activation fit results.
Obtained fitting parameters for {\it s}-shell and {\it p}-shell excitons are shown in (c) and (d), respectively.
}
\label{fig6}
\end{figure}

To gain insight into the phonon interactions, we measured the temperature dependence of the dephasing rates and population decay rates. 
Figure 6(a) shows the temperature dependence for the dephasing rates of ZPLs of {\it s}-shell and {\it p}-shell excitons. For both shells, the dephasing rates rapidly increase near 40 K.
At the lowest temperature, on the other hand, the dephasing rate for {\it p}-shell excitons is significantly larger than {\it s}-shell excitons.
For {\it s}-shell excitons, the absence of a temperature dependence up to 100 K for population decay rates has also been observed in time-resolved photo-luminescence experiments \cite{Bardot2005}.
It is found in this work that the population lifetime of {\it p}-shell excitons is also robust against temperature, although the {\it p}-shell excitons are energetically closer to QWs and spatially less confined than {\it s}-shell excitons.
From these observations, we conclude that net population for each shell is maintained even though population transfers might occur within multiple electron and hole states in {\it p}-shell excitons by emissions or absorptions of LA phonons.

Because the population dynamics are not affected by changes in the temperature, as shown in Fig. 6(b), the linewidth thermal broadening is associated with elastic, pure dephasing mechanisms. Figures 6(c) and 6(d) show the temperature-dependent pure dephasing rates for {\it s}-shell and {\it p}-shell excitons, respectively. To obtain quantitative insight, we fit data with a single-phonon activation model given by
\begin{equation}
\gamma^{\ast} \left(T_L \right) = \gamma_0 + \gamma_1 \frac{1}{\exp \left( \frac{E_1}{k_{\text{B}}T_L} \right) + 1},
\end{equation}
where $k_{\text{B}}$ is the Boltzmann constant, $\gamma_0$ is the pure dephasing rate at $T_L$ = 0 K, $\gamma_1$ represents the exciton-phonon coupling strength, and $E_1$ is activation energy of the single phonon mode. The black solid lines are fitted results, and obtained parameters are indicated in each panel.

At the lowest temperature, $\gamma_0 \simeq 0$ for {\ s}-shell excitons indicates that the dephasing is limited by the population decay rate, which is in good agreement with the previous report \cite{Langbein2004}.
For the {\it p}-shell, $\gamma_0$ = 10 $\mu$eV clearly shows the presence of a substantial amount of pure dephasing even at low temperature.
This result contradicts a previous report relying on FWM measurements quantitatively \cite{Borri2006}, where two time scales for dephasing are measured, {\it i.e.} a long dephasing time of 1.1 ns ($\gamma_0$ = 0.6 $\mu$eV as HWHM) and short one of 12 ps. In that work, the dephasing was explained as a result of the characteristic energy structure of {\it p}-shell excitons \cite{Muljarov2006_PSS}, where the lower and upper bright excitons were energetically split by as much as 2.9 meV and two additional dark exciton states lie near the upper bright exciton.
As a consequence, the upper bright excitons suffered from significant dephasing due to interactions with dark excitons, while the lower excitons were nearly free from dark excitons and exhibit long dephasing time.

\begin{figure}[!b]
\includegraphics{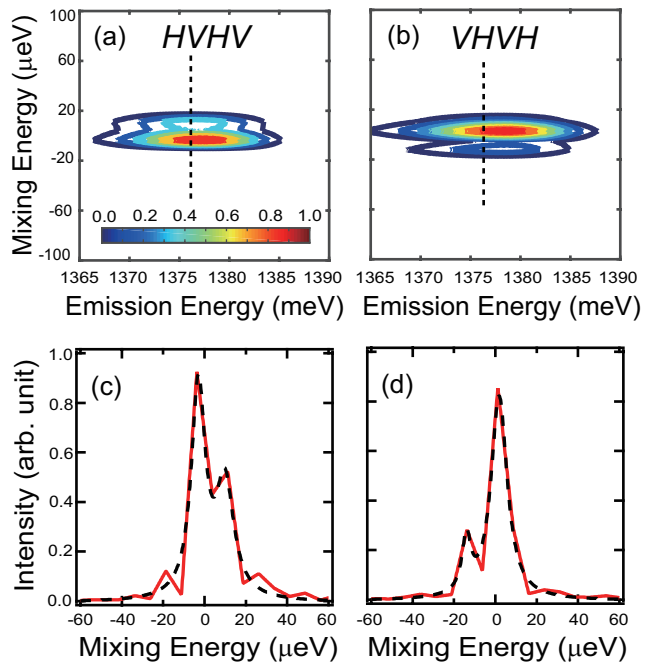}
\caption{(Color online) Rephasing zero-quantum amplitude spectra for p-shell excitons with the polarization sequences of (a) $HVHV$ and (b) $VHVH$.
(c), (d) Intensity profiles of vertical lines shown as dashed lines in (a) and (b).
Red solid lines are data and black dashed lines are double-Lorentzian fits.
}
\label{fig7}
\end{figure}

The energy level structure of the QDs studied in our work is different.
The sample used in the previous report \cite{Borri2006} had a much larger confinement energy of 332 meV leading to an energy splitting of 2.9 meV, whereas the confinement energy is 100 meV in our sample.
In order to determine the energy splitting between the two bright {\it p}-shell excitons, we perform rephasing zero-quantum scans in a cross-linearly polarized configuration ({\it HVHV} and {\it VHVH}), from which the energy splitting between bright $H$- and $V$-excitons is revealed by accessing the non-radiative coherence between the two bright excitons \cite{Moody_SSC}.
Figures 7(a) and 7(b) show the 2D rephasing zero-quantum amplitude spectra for {\it p}-shell excitons with polarization sequences of {\it HVHV} and {\it VHVH}, respectively.
Additional peaks above and below the main peaks at the mixing energy of 12 $\mu$eV are clearly observed in Figs. 7(a) and 7(b), respectively, as a consequence of a fine structure splitting of {\it H}- and {\it V}-excitons for {\it p}-shell excitons.
To clearly show this splitting, the lineshapes along the zero-quantum axis are shown in Figs. 7(c) and 7(d), respectively.
The red solid lines represent data whereas the black dashed lines are double-Lorentzian fits.
The energy splitting amounts to 14 $\pm$ 3(15 $\pm$ 3) $\mu$eV from {\it HVHV} ({\it VHVH}), which is similar to the splitting observed for {\it s}-shell excitons~\cite{Moody_SSC}.
The similarity between the bright-state splitting of both shells indicates that, like the {\it s}-shell excitons, the {\it p}-shell excitons are also not affected by interactions between the bright and dark states due to their large energy difference.
We therefore attribute the low-temperature pure dephasing to scattering between the multiple, nearly-degenerate bright states composing the ${\it p}$-shell excitons.
This interpretation is consistent with a previous report on the dephasing rates of excited QD states, which showed that $\gamma_0$ for {\it p}-shell excitons is due to population transfer between inter-level hole states mediated by LA phonon emission \cite{Htoon2001}.
In our QDs, the {\it p}-shell exciton can scatter between the nearly degenerate transitions, which preserves the net population in the shell but gives rise to additional dephasing.
The contrasting results of $\gamma_0$ for {\it s}- and {\it p} - shell excitons also suggest that quasi-elastic scattering becomes possible within the {\it p}-shell excitons without change of the spin orientation while for scattering in {\it s}-shell excitons spin-flip is strongly suppressed.

The phonon dephasing mechanism can be inferred by the activation energies, which are 22 and 24 meV for {\it s}-shell and {\it p}-shell excitons, respectively.
These values are close to energy separation between {\it s}-shell and {\it p}-shell excitons.
Systematic confinement-dependent measurements for {\it s}-shell dephasing rates, however, clearly show the independence of activation energy on confinement energy \cite{Borri2005}.
Because inter-shell energy separations are determined by confinement energy, this previous study strongly suggests that real energy transitions between shells are irrelevant.
We also rule out real phonon-mediated transitions between shells, since these would give rise to additional cross-peaks in the spectra\cite{Moody2011_2}, which are not observed.
Since the respective LO phonon energies of 30 meV and 32 meV for the InAs wetting layer and QDs \cite{Heitz1996} are similar to the obtained activation energies, another possible mechanism consists of elastic virtual LO phonon transitions.
However, this mechanism was theoretically proven to not result in dephasing in the finite number of discrete excitonic states\cite{ Muljarov2007, Muljarov2006}.
Thus, we conclude that the relevant mechanism to account for our results is the pure dephasing caused by virtual transitions via quadratic coupling to acoustic phonons \cite{Muljarov2004, Zibik2008, Grange2009}.
It is interesting to note that the obtained value of activation energy for {\it p}-shell excitons is similar to that of the {\it s}-shell.
This quantitative match of activation energies indicates that the dephasing of {\it p}-shell excitons is also affected more or less by the same mechanism as the {\it s}-shell excitons, namely quadratic couplings to acoustic phonons \cite{Muljarov2004}.

The larger value of $\gamma_1$ for the {\it p}-shell as compared to the {\it s}-shell directly shows the stronger exciton-phonon couplings for {\it p}-shell.
Along the theory describing quadratic couplings, the electron({\it e})/hole({\it h})-phonon coupling is quantified by the matrix element for deformation potential interaction with longitudinal acoustic phonons, explicitly expressed by \cite{Muljarov2004}
\begin{equation}
M_{a, \bm{k}}^{i,j}  = \sqrt{\frac{\hbar \omega_k}{2 \rho c_s^2 V}} D_a \int d\bm{r} \psi^{\ast}_{ia} \left(\bm{r} \right) e^{i \bm{k} \cdot \bm{r}} \psi_{ja} \left(\bm{r} \right)
\end{equation}
where $a =e,h$, $\bm{k}$ is the phonon wave vector, $i, j$ are the electron/hole states for involved transitions, $\rho$ is the mass density, $c_s$ is the sound velocity, $V$ is the phonon renormalization volume, and  $D_a$ is deformation potential.
$\psi_{ia}$ is the confinement wave function, and the different values for matrix elements between {\it s}-shell and {\it p}-shell come from the integration in Eq. (2).
Due to the two closely lying electron and hole energy levels of which {\it p}-shell excitons consist, an increased number of combinations for $i$ and $j$ states is possible, which results in a larger phonon contribution to pure dephasing rates than in the case of {\it s}-shell excitons.

\section{summary}
We have studied exciton-phonon interactions for {\it p}-shell excitons of InAs self-assembled quantum dots and compared them to those obtained for {\it s}-shell excitons by using two dimensional coherent spectroscopy.
Systematic measurements for dephasing rates of {\it p}-shell excitons have revealed pure dephasing mechanisms to be most important.
Larger values of pure dephasing rates at the low temperature ($\gamma_0$) and stronger exciton-phonon coupling ($\gamma_1$) stems from multiple closely lying electron and hole levels comprising {\it p}-shell excitons.
%We anticipate our result can provide quantitative information on the engineering using {\it p}-shell excitons as a platform of next generation devices.

\begin{acknowledgements}
The work at JILA \& University of Michigan was primarily supported by the Chemical Sciences, Geosciences, and Energy Biosciences Division, Office of Basic Energy Science, Office of Science, U.S. Department of Energy under Awards No. DE-FG02-02ER15346 and DE-SC0015782.
T. S. acknowledges support by Japan Society for the Promotion of Science (JSPS).
A.D.W. gratefully acknowledges support of Mercur  Pr-2013-0001, DFG-TRR160 Z1,  BMBF - Q.com-H  16KIS0109, and the DFH/UFA  CDFA-05-06.
\end{acknowledgements}

\end{document}